\documentclass[a4paper,11pt]{article}
\usepackage{pos}
\usepackage{amsmath,bm,color,fancyhdr,hyperref,lastpage,units}
\usepackage{graphicx}
\usepackage{enumitem}
\usepackage{kantlipsum}
\usepackage{array}
\usepackage[english]{babel}

\newcommand{\Tr}{\mathrm{Tr}}
\newcommand{\G}{\Gamma}
\newcommand{\Gd}{\Gamma^{\dagger}}
\renewcommand{\Re}{\mbox{\textrm{Re}}}

\title{Meson screening mass at finite chemical potential
}
\ShortTitle{Meson screening mass at finite chemical potential
}

\author*[a]{Rishabh Thakkar}
\author[a]{Prasad Hegde}

\affiliation[a]{Indian Institute of Science, Bangalore}

\emailAdd{rishabht@iisc.ac.in, prasadhegde@iisc.ac.in}

\abstract{Knowledge of the screening masses at finite chemical potential can provide insight into the nature of the QCD phase diagram. However, lattice studies at finite chemical potential suffer from the well-known issue of the sign problem, which has made the calculation of observables such as screening correlators and screening masses at finite chemical potential quite challenging. One way to proceed is by expanding the observable in a Taylor series in the chemical potential and hence calculating the finite-density corrections to the observable. In this talk, we will use this approach to calculate the screening mass of the pseudoscalar meson at finite temperatures and chemical potential by expanding the screening correlator in a Taylor series in the chemical potential. We will present our results for the second derivative of the screening mass w.r.t. the chemical potential. Our calculation was done on $64^3 \times 8$ lattices generated using the (2+1) HISQ/tree action.}

\FullConference{%
The 39th International Symposium on Lattice Field Theory,\\
8th-13th August, 2022,\\
Rheinische Friedrich-Wilhelms-Universität Bonn, Bonn, Germany
}

\begin{document}
\maketitle

\section{Introduction}
\label{sec:introduction}
QCD undergoes a phase transition from hadronic degrees of freedom at low temperature to the deconfined quarks and gluons at high-temperature \cite{aoki2006order,bazavov2019chiral}. At zero density, this transition is a crossover with the transition being continuous. A lattice analysis marks this pseudo-critical temperature at $T_{\text{pc}}=156.5\pm1.5$ MeV \cite{bazavov2019chiral}. Upon introducing non-zero density, this transition is expected to approach a second-order critical point beyond which a first-order transition marks the phase transition. The location of this critical point is still elusive. To locate this critical point it is necessary to understand the behavior of QCD at finite density and temperature. However, this becomes difficult for lattice simulations because of the complex fermion determinant at a finite chemical potential. This complex determinant poses a challenge in simulating the system on the lattice because of an oscillating weight inside the partition function integral. Various numerical methods are used to resolve this challenge for small chemical potential \cite{nagata2022finite}. In this work, the required observables are expanded in the Taylor series of the chemical potential and then  calculated on lattices generated at zero chemical potential . This gives their response to the chemical potential as a series expansion.

The nature of hadronic excitations has phenomenological importance for understanding the interaction between the quarks and gluons. We work with spatial correlation functions of mesonic operators to obtain the mesonic excitations. These spatial correlators, also called screening correlators, are obtained by propagating a quark and anti-quark in the spatial direction. They decay exponentially at a large distance and the decay constant is called the screening mass and is the inverse of the screening length. On the approach to the critical point, this screening length diverges as the long-distance correlation becomes significant requiring the screening mass to vanish at the critical point.

There has been a previous study to calculate the corrections to the screening mass at finite density previously \cite{Pushkina:2004wa} using the Taylor expansion. Our work extends that study by allowing the screening mass to take complex values. The motivation for this consideration is the free theory analytical expression for the screening correlator with an oscillatory behavior suggesting a complex screening mass \cite{vepsalainen2007mesonic}. The analysis carried out in this paper focuses only on the pseudoscalar channel meson for the isoscalar chemical potential $\mu_S$ where 
\begin{equation}
    \mu_S=\mu_u=\mu_d.
\end{equation}
with $u$ and $d$ representing the up and down quarks.

\section{Screening correlator and screening mass}
\label{sec:formalism}
\par On lattice, mesonic correlation functions are two-point functions represented by 
\begin{flalign}
C(n,\mu_i,\mu_j)\,\,=&\,\,\left\langle O_\Gamma(n)\Bar{O}_\Gamma(0)\right\rangle\notag\\
=&\,\,\langle G(n,\mu_i,\mu_j)\rangle \notag\\ \,\,=& \,\,\left\langle\Tr\left[P(\mu_i)_{n,0}\Gamma P(\mu_j)_{0,n}\Gamma^\dagger\right]\right\rangle
\label{eq:corr_finite}
\end{flalign}
where $O$ is the meson operator given by $O=\Bar{\psi}_i\Gamma\psi_j$ with $i,j$ corresponding to the flavor indices of the quark field $\psi$, $\Gamma$ is the Dirac spin matrix corresponding to the spin of the meson, $G(n)$ is the meson propagator propagating from the origin to Euclidean space-time site $n$ on the lattice and  $P(\mu_i)_{n,0}$ is the quark propagator which is the inverse of the Dirac operator ${\mathcal{M}}(\mu_i)_{n,0}$
\begin{equation}
    P(\mu_i)_{n0}={\mathcal{M}}^{-1}(\mu_i)_{n0}
\end{equation}
\par Using the modified $\gamma_5-$hermiticity property of the finite $\mu$ lattice Dirac operator $\mathcal{M}$, {\it i.e.} $\mathcal{M}(\mu_i)_{0n}=\gamma_5\mathcal{M}(-\mu_i)^\dagger_{n0}\gamma_5$, we can obtain the conjugation property of the meson propagator
\begin{eqnarray}
G(n,\mu_i,\mu_j)& =& \Tr \left[ P(\mu_i)_{n 0}\G
\gamma_5P(-\mu_j)_{n 0}^{\dagger}\gamma_5 \Gd \right]\\
G(n,\mu_i,\mu_j)^*& =& \Tr \left[ \G
\gamma_5P(-\mu_j)_{n 0}\gamma_5 \Gd P(\mu_i)_{n 0}^{\dagger} \right]\notag\\
&=&\Tr \left[ 
\gamma_5P(-\mu_i)_{0 n}\gamma_5 \G P(\mu_j)_{0 n}^{\dagger}\Gd \right]\notag\\
&=&\Tr \left[ 
P(-\mu_i)_{0 n} \G \gamma_5 P(\mu_j)_{0 n}^{\dagger}\gamma_5\Gd \right]\notag \\
\implies G(n,\mu_i,\mu_j)^*&=& G(n,-\mu_i,-\mu_j)
\label{eq:G_exp_conj}
\end{eqnarray}

\par At zero temperature, the asymptotic large Euclidean time behavior of the correlator yields the ground state excitation. At finite temperatures, the temporal extent of the lattice is constrained by the temperature of the system, $N_\tau=1/T$.
However, there are no such constraints in the spatial directions making it easy to analyze the screening correlator at large distances. These screening correlators are obtained by summing over the $x,y$, and $t$ directions
\begin{flalign}
C(n_z,\mu_i,\mu_j)\,\,=&\,\,\sum_{x,y,t}\langle G(n,\mu_i,\mu_j)\rangle \notag\\ \,\,=& \,\,\sum_{x,y,t}\left\langle\Tr\left[P(\mu_i)_{n,0}\Gamma P(\mu_j)_{0,n}\Gamma^\dagger\right]\right\rangle
\label{eq:corr_finite}
\end{flalign}
\par For $\mu=0$, these correlators decay exponentially and for the periodic boundary condition of lattice, we get
\begin{eqnarray}
C(n_z)&=&\sum_iA_i'\left(e^{-M_i\left(n_z-\frac{N_\sigma}{2}\right)}+e^{M_i\left(n_z-\frac{N_\sigma}{2}\right)}\right)\notag\\
&=&\sum_iA_i\cosh\left[M_i\left(n_z-\frac{N_\sigma}{2}\right)\right]
\end{eqnarray}
where $i$ is the sum over excitation states, $M$ is the screening mass in lattice units and $N_\sigma$ is the spatial extent of the lattice.
\par The staggered quarks have four spin and four taste indices giving sixteen mesons for each meson channel specified by $\Gamma=\Gamma_D\times\Gamma_T$ with $\Gamma_D$ and $\Gamma_T$ being the Dirac Gamma matrices for spin and taste structures, respectively. We will limit our analysis only to local meson operators with $\Gamma_D$ = $\Gamma_T$ = $\Gamma$. With this, the local meson operators reduce to a product of phase factor $\phi(n)$ and bilinear of staggered quarks $\chi(n)$, $M(n)=\phi(n)\Bar{\chi}_i(n)\chi_j(n)$\cite{altmeyer1993hadron}.
For a constant separation between the source and the sink for the staggered correlator, the contribution from two sets of mesons with the same spin but with opposite parities are summed which are given by
\begin{eqnarray}
C(n_z)&=&\sum_iA_i^{(-)}\cosh\left[M_i^{(-)}\left(n_z-\frac{N_\sigma}{2}\right)\right]-(-1)^{n_z} A_i^{(+)}\cosh\left[M_i^{(+)}\left(n_z-\frac{N_\sigma}{2}\right)\right]
\label{eq:stag_corr}
\end{eqnarray}
The local meson operator for the pseudoscalar channel is given by $\Gamma^{(-)}=\gamma_5$ and $\Gamma^{(+)}=\gamma_3$ corresponding to $\phi(n)=1$ \cite{cheng2011meson}. The oscillating $(+)$ channel for pseudoscalar meson is conserved which doesn't excite any states from the vacuum \cite{altmeyer1993hadron} and thus has contribution only from the non-oscillating $(-)$ channel.

\section{Isoscalar chemical potential response for the free theory correlator}
\par For finite $\mu$, analytical expression exists only for the free theory case \cite{vepsalainen2007mesonic}. When simplified for isoscalar chemical potential, the free theory correlator equation becomes 
\begin{eqnarray}
\displaystyle C(z,\mu_S)&=&\displaystyle \frac{3T^2}{2 z}e^{- 2  \pi T z}\left[\left(1 + \frac{1}{2  \pi T z}\right)\cos(2\mu_S z) + \frac{\mu_S}{\pi T}\sin(2\mu_S z)\right] +\mathcal{O}(e^{-4\pi Tz})\notag\\
&=&\displaystyle \frac{3T^2}{2 z}\Re\left[ e^{- 2  \pi T z + i\, 2\mu_S z }\left(\left(1 + \frac{1}{2  \pi T z}\right) - i\frac{\mu_S}{\pi T}\right) \right] +\mathcal{O}(e^{-4\pi Tz})
\label{eq:corr_isoscalar_chem}
\end{eqnarray}
While the above correlator itself is real, it has periodic oscillations due to the screening mass and the amplitude having a complex value given by 
\begin{eqnarray}
M_{scr}^{free}&=&2  \pi T  + i\, 2\mu_S 
\label{eq:mass_free}\\
A^{free}&=&\frac{3T^2}{2 z}\left(\left(1 + \frac{1}{2  \pi T z}\right) - i\frac{\mu_S}{\pi T}\right)
\end{eqnarray} 
The imaginary part of the screening mass and amplitude depend linearly on the chemical potential while the real part is independent of the chemical potential.
\par To the leading term, in the limit $\mu_{S}\rightarrow 0$, we get 
\begin{eqnarray}
\displaystyle C(z,0)=\displaystyle \frac{3T^2}{2 z}e^{- 2  \pi T z}\left(1 + \frac{1}{2  \pi T z}\right)
\label{eq:corr_free_theory}
\end{eqnarray}
which is the expected exponential fall-off. To observe the response of the correlator around $\mu_S=0$, derivatives of correlators are obtained.
Taking the derivative of the \eqref{eq:corr_isoscalar_chem} with the isoscalar chemical potential, the odd derivatives vanish. The first two non-zero derivatives at $\mu_S=0$ are
\begin{eqnarray}
C''(z)=&- 6\displaystyle T^{2} e^{- 2 \pi z T}\displaystyle  \left(z \left(1 + \frac{1}{2\pi z T}\right) - \frac{1}{\pi T}\right) &\\
C''''(z)=&\displaystyle 12 z^{2}  T^{2} e^{- 2 \pi z T}\left(2z \left(1 + \frac{1}{2\pi z T}\right) - \frac{4}{\pi T}\right)&
\label{eq:free_corr_deriv}
\end{eqnarray}
To get rid of the contribution of the exponential decay, we define $\Gamma$ and $\Sigma$ by dividing the above equations by the free theory correlator at $\mu_S=0$ 
\begin{eqnarray}
	\Gamma_{Free}(z)\equiv\frac{C''(z)}{C(z)} =& - 4z\displaystyle  \left(z  - \frac{1}{\pi T\left(1 + \frac{1}{2\pi z T}\right)}\right)&
		\label{eq:free_corr_by_corr_isoscalar2}\\
	\Sigma_{Free}(z)\equiv\frac{C''''(z)}{C(z)}=&\displaystyle 16 z^{3} \left(z  - \frac{2}{\pi T\left(1 + \frac{1}{2\pi z T}\right)}\right)&
	\label{eq:free_corr_by_corr_isoscalar4}
\end{eqnarray}
Thus, at a large distance, we obtain $\Gamma_{free}$ and $\Sigma_{free}$ which are quadratic and quartic in $z$ respectively upto $\mathcal{O}(1/z)$ corrections.

\section{Isoscalar chemical potential response for correlator at finite temperature}
\par The correlators and screening masses change analytically with temperature. Thus, at very high temperatures and finite isoscalar chemical potential, we expect the correlator to have a similar behavior as the free theory correlator given in \eqref{eq:corr_isoscalar_chem}. Considering the amplitude and screening mass to be a function of the isoscalar chemical potential and having  a contribution from only the ground state, write them as
\begin{eqnarray}
C(z;\mu_S)& = & \Re\left[\left(A_R(\mu_S)-iA_I(\mu_S)\right)e^{-z(M_R(\mu_S)+iM_I(\mu_S)} \right] \notag\\ & = &e^{-zM_R(\mu_S)}\left\{A_R(\mu_S)\cos\big(zM_I(\mu_S)\big) + A_I(\mu_S)\sin\big(zM_I(\mu_S)\big)\right\}
\label{eq:corr_complex}
\end{eqnarray}
where the screening mass is $M_{scr}(\mu_S)=M_R(\mu_S)+iM_I(\mu_S)$ and the amplitude is $A(\mu_S)=A_R(\mu_S)-iA_I(\mu_S)$.
Using $\mu_i=\mu_j=\mu_S$ in \eqref{eq:G_exp_conj}, we get constraint on the correlator 
\begin{eqnarray}
C(z; - \mu_S) = C^*(z; \mu_S)
\label{eq:constraint}
\end{eqnarray}
For this relation to be satisfied, we must have $M_R (-\mu_S) = M_R (\mu_S)$. For free theory we have $M_I (-\mu_S) = -M_I (\mu_S)$ for free theory and we expect the same behavior at finite temperature.
This requires $A_{R} (-\mu_S) =  A_{R} (\mu_S)$ and $A_{I} (-\mu_S) = - A_{I} (\mu_S)$ to satisfy \eqref{eq:constraint}.  Thus, the real and imaginary parts of the screening mass
and amplitude are even and odd functions of $\mu_S$ respectively having even and odd power of $\mu_S$ in the Taylor expansion. With this constraint, the correlator \eqref{eq:corr_complex} can be expanded in terms of $\mu_S$
where all the derivatives of $A$ and $M$ are obtained at $\mu_S=0$. 
Collecting the terms with second and fourth powers of $\mu_S$, and dividing them by the correlator, we obtain
\begin{eqnarray}
\Gamma(z)\equiv\left.\frac{d^2C}{Cd\mu_S^2}\right|_{\mu_S=0}=\left.\frac{C''}{C}\right|_{\mu_S=0} &= &\frac{A_R''}{A_R} 
+ z\left[2\frac{A_I'}{A_R}M_I' - M_R''\right]
- z^2(M_I')^2.
\label{eq:Gamma}\\
&=&\alpha_2z^2+\alpha_1z+\alpha_0\\
\Sigma(z)\equiv\left.\frac{d^4C}{Cd\mu_S^4}\right|_{\mu_S=0}=\left.\frac{C''''}{C}\right|_{\mu_S=0}&= &\frac{A_R''''}{A_R} + z(4\frac{A_I'}{A_R}M_I'''+4\frac{A_I'''}{A_R}M_I'-M_R''''-6M_R'' \frac{A_R''}{A_R}) \notag\\&&+z^2(3M_R''^2-12\frac{A_I'}{A_R}M_I'M_R''-4M_I'M_I'''-6 M_I'^2\frac{A_R''}{A_R})
\notag\\&&+z^3(6M_R''M_I'^2-4\frac{A_I'}{A_R}M_I'^3)+z^4(M_I'^4)
\label{eq:Sigma}\\
&=&\beta_4z^4+\beta_3z^3+\beta_2z^2+\beta_1z+\beta_0
\end{eqnarray}
Similar to the free theory expression 	\eqref{eq:free_corr_by_corr_isoscalar2} and \eqref{eq:free_corr_by_corr_isoscalar4},
$\Gamma(z)$ is quadratic in $z$ and $\Sigma(z)$ is quartic in $z$.
Using the equations \eqref{eq:Gamma} and \eqref{eq:Sigma}, we get $M_I'$ and $M_R''$ as
\begin{eqnarray}
M_I'=-\alpha_2^{1/2}=\beta_4^{1/4}\label{eq:mi1}\\
M_R''=\frac{1}{4}\left(2\alpha_1-\frac{\beta_3}{\alpha_2}\right)
\label{eq:mr2}
\end{eqnarray}
\par Although we have considered only the ground state contribution to the correlator, there is a significant contribution from the excited states to the correlators at finite spatial distances \cite{bazavov2019meson} and we need to include their contributions into the expressions of $C, \Gamma$ and $\Sigma$. Below we consider the contribution of the first excited state into their equations with labels 0 and 1 corresponding to the ground and first excited states respectively.
\begin{eqnarray}
C(z)&=&A_0\exp^{-M_0z} + A_1\exp^{-M_1z} = A_0\exp^{-M_0z}\left[1 + \frac{A_1}{A_0}\exp^{-(\Delta M)z}\right]\\
\Gamma(z)&=&\frac{\left(\alpha_{02}z^2+\alpha_{01}z+\alpha_{00}\right) + \frac{A_1}{A_0}\exp^{-(\Delta M)z}\left(\alpha_{12}z^2+\alpha_{11}z+\alpha_{10}\right)}{1 + \frac{A_1}{A_0}\exp^{-(\Delta M)z}}\notag\\
&\simeq&\frac{\left(\alpha_{02}z^2+\alpha_{01}z+\alpha_{00}\right) }{1 + \frac{A_1}{A_0}\exp^{-(\Delta M)z}}\label{eq:Gamma_excited}\\
\Sigma(z)&=&\frac{\left(\beta_{04}z^4+\beta_{03}z^3+\beta_{02}z^2+\beta_{01}z+\beta_{00}\right) + \frac{A_1}{A_0}\exp^{-(\Delta M)z}\left(\beta_{14}z^4+\beta_{13}z^3+\beta_{12}z^2+\beta_{11}z+\beta_{10}\right)}{1 + \frac{A_1}{A_0}\exp^{-(\Delta M)z}}\notag\\
&\simeq&\frac{\left(\beta_{04}z^4+\beta_{03}z^3+\beta_{02}z^2+\beta_{01}z+\beta_{00}\right) }{1 + \frac{A_1}{A_0}\exp^{-(\Delta M)z}}
\label{eq:Sigma_excited}
\end{eqnarray}
Thus, when including the contribution for the first excited states, both $\Gamma$ and $\Sigma$ are quadratic and quartic with an exponential decaying denominator reaching the value of the ground state coefficient asymptotically.

\section{Lattice setup}
All the numerical data in this work used the Bielefeld GPU code \cite{altenkort2021hotqcd} to generate the lattices. They were constructed by simulating staggered fermion operators using (2+1) flavor HISQ action gauge field ensembles. The strange mass for the configurations was tuned to the physical mass by tuning the mass of $\eta_{\Bar{s}s}$ meson $M_{\eta_{\Bar{s}s}}=686 $ MeV \cite{bazavov2019meson} and ratio $m_s/m_l=20$ is kept constant corresponding to the pion mass 160 MeV. The lattice scale is set using the kaon decay constant $f_K=156.1/\sqrt{2}$ MeV. The configurations were generated using the
leapfrog evolution with molecular dynamics step size  0.2 and trajectory length of 5 steps keeping the acceptance rate between 65\% to 80\%. The meson correlators are measured on every $10^{th}$ configuration.
\par The free theory analysis was performed for lattice volume $80^3\times 8$. The finite temperature analysis was done for $N_\sigma=64$ with $N_\tau=8$. The corresponding configurations and quark masses are presented in table \ref{tab:conf_list}. 1000 random source vectors were used for estimating traces on each configuration. 8 point sources were used on each configuration to measure the correlator-like operator.
\begin{table}[h!]
	\centering
	\begin{tabular}{ |c|c|c|c|c|c| } 
		\hline
		$N_\sigma$ & $\beta$ & T[GeV]& $m_l$ & $m_s$ & configurations \\ 
		\hline
		64 & 9.670 & 2.90 & 0.0001399 & 0.002798& 6000\\ 
		64 & 9.360 & 2.24 & 0.00018455& 0.003691& 6000	\\
		\hline
	\end{tabular}
	\caption{The list of configurations used for the finite temperature. All the configurations used here have $N_\tau=8$.}
	\label{tab:conf_list}
\end{table}
\section{Lattice results}
\begin{figure}[t]
\centering
\includegraphics[width=0.45\textwidth]{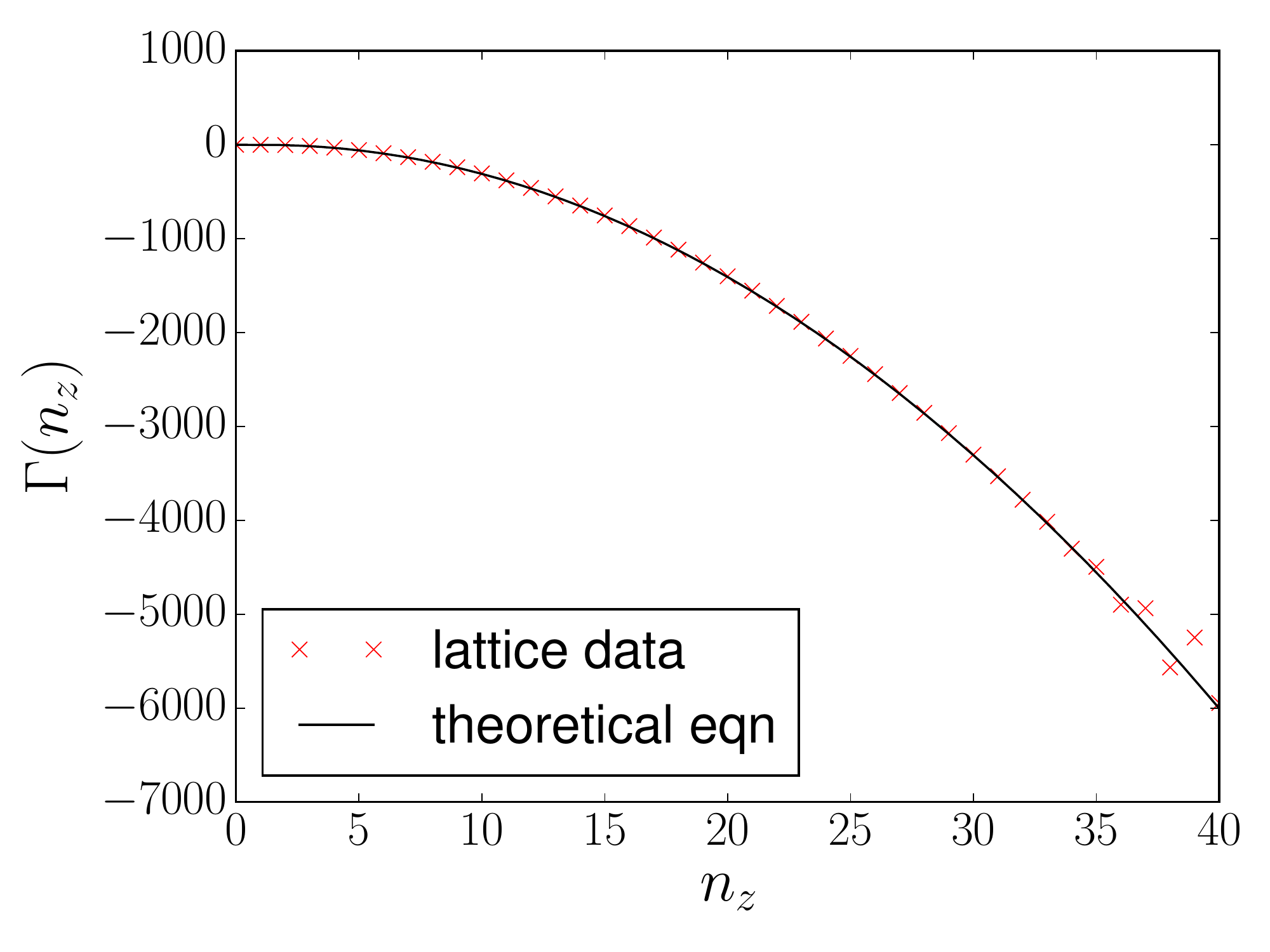}
\includegraphics[width=0.45\textwidth]{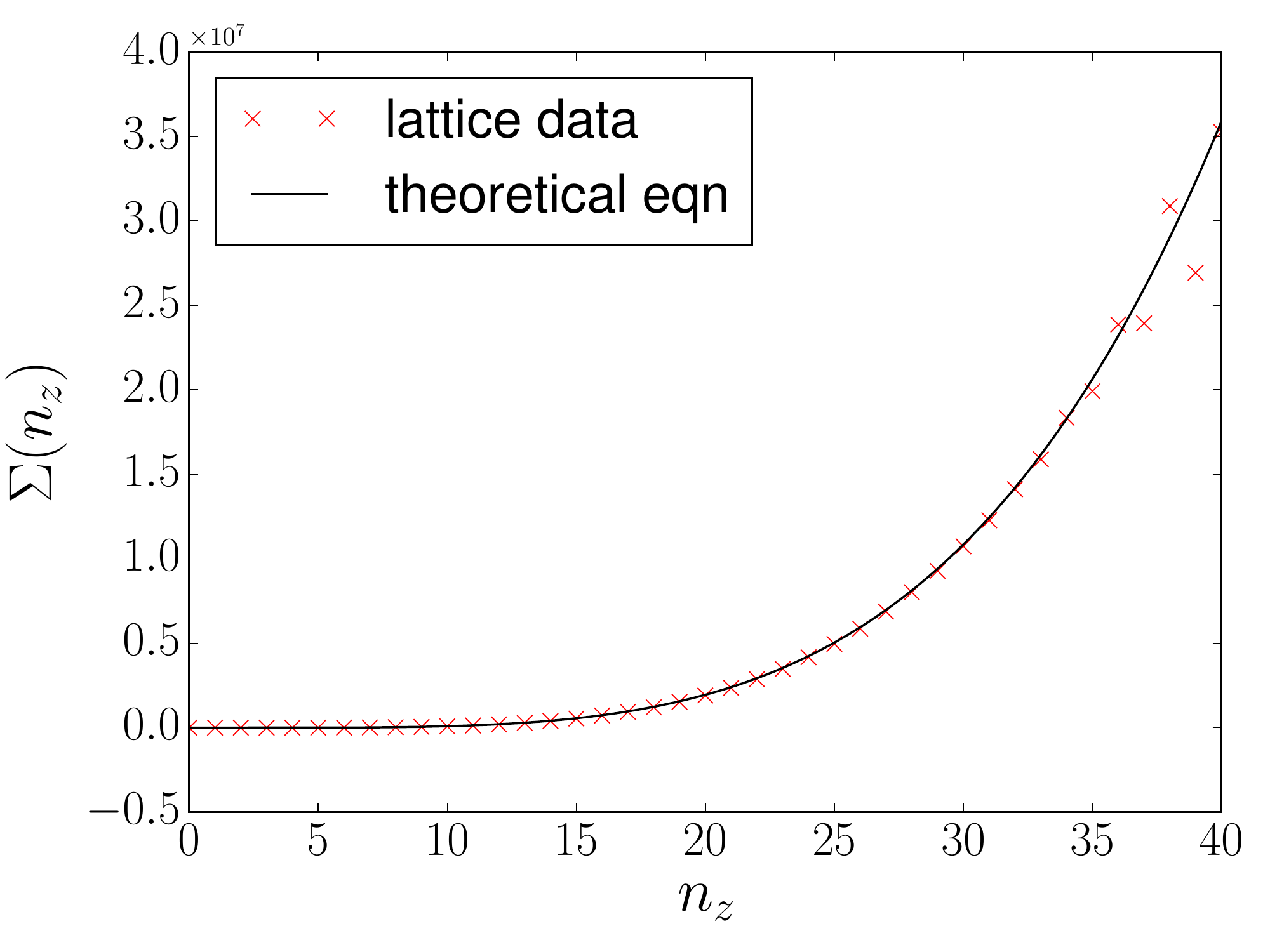}
	\caption{ (left) $\Gamma(n_z)$ and (right) $\Sigma(n_z)$ plotted for free theory with against $n_z$ calculated on lattice with volume $80^3 \times 8$ for pseudoscalar meson. The solid curves in both figures are theoretical equations given in \eqref{eq:free_corr_by_corr_isoscalar2} and \eqref{eq:free_corr_by_corr_isoscalar4}.}
	\label{fig:free_corr_by_corr}
\end{figure}
\begin{figure}[b]
\centering
\includegraphics[width=0.45\textwidth]{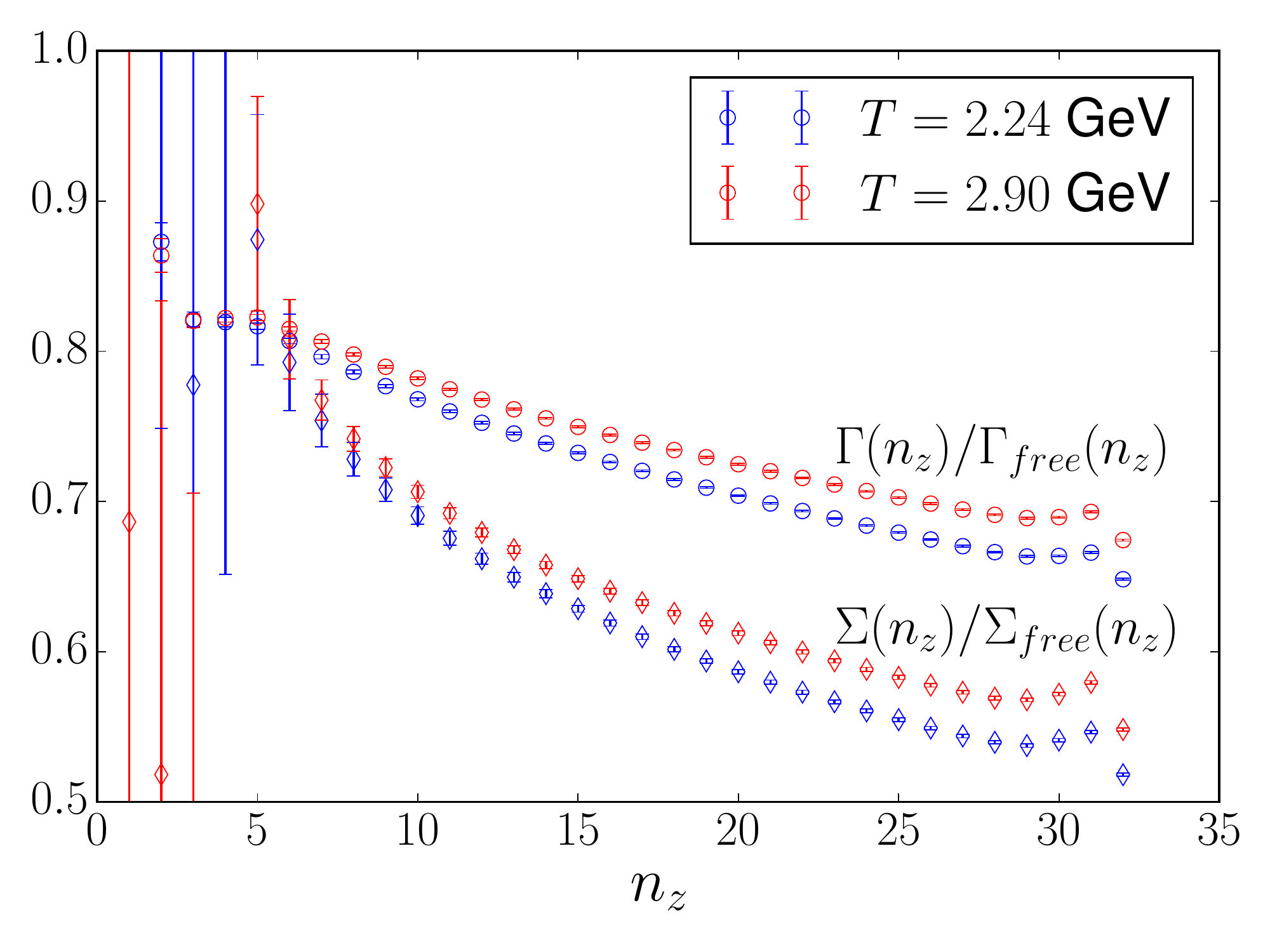}
\includegraphics[width=0.45\textwidth]{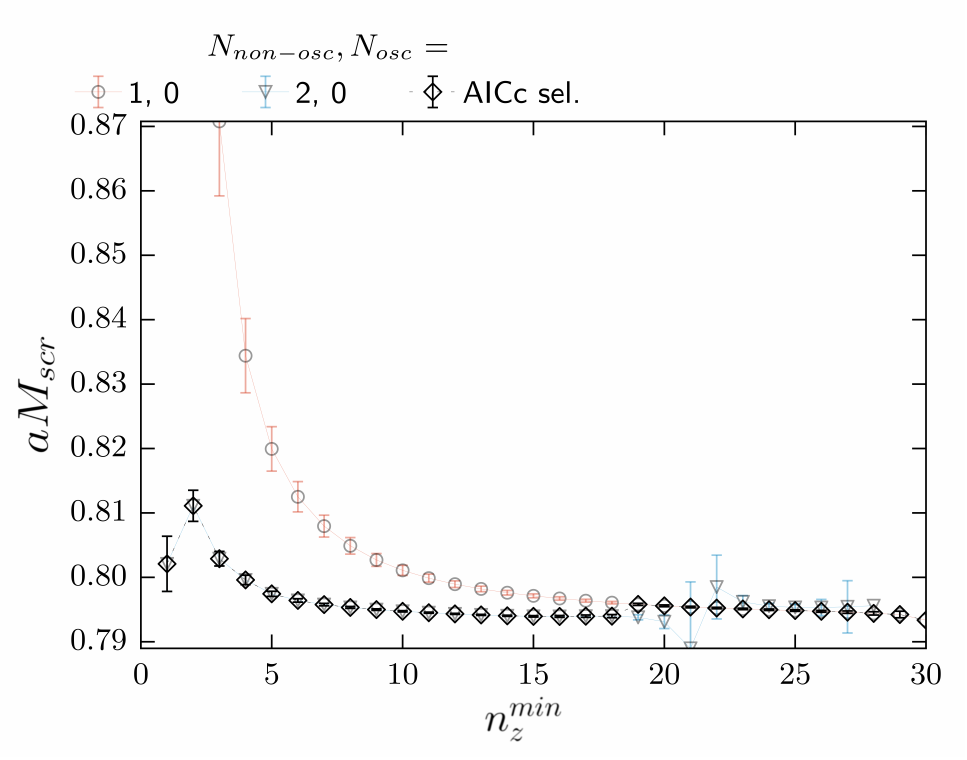}
	\caption{ (left) $\Gamma/\Gamma_{free}$ and $\Sigma/\Sigma_{free}$ plotted against $n_z$ for $T=2.24$ GeV and $T=2.90$ GeV. (right) Screening mass of the  $T=2.90$ GeV correlator obtained by fitting the $\mu_S=0$ correlator using ground state (1,0) and first excited state (2,0) ansatz. The best fitting function is chosen using Akaike criteria AICc \cite{bazavov2019meson}. The data points in both the figures are obtained using lattices with dimension $64^3\times8$ using point source and $m_s/m_l=20$. }
	\label{fig:finite_temp1}
\end{figure}
\par The lattice expressions for derivatives of the correlator for the staggered fermions are given in \cite{Pushkina:2004wa,rishabh} which are obtained by taking $\mu_S$ derivatives of the meson propagator $G$ and the staggered fermionic determinant. Using these we obtained the results discussed below.
\par In figure \ref{fig:free_corr_by_corr}, we plot the lattice results for the free theory comparing $\Gamma_{free}$ and $\Sigma_{free}$ obtained on the lattice with the theoretical expression \eqref{eq:free_corr_by_corr_isoscalar2} and \eqref{eq:free_corr_by_corr_isoscalar4} respectively. The good agreement between the theoretical expressions and our lattice data provides support for our proposed ansatz \eqref{eq:corr_complex}.  
\par For finite temperature, we first look at the effect of temperature on $\Gamma$ and $\Sigma$. To see this, we plot $\Gamma/\Gamma_{free}$ and $\Sigma/\Sigma_{free}$ for two temperatures in figure \ref{fig:finite_temp1} (left). As we go to higher temperatures, we expect these expressions to approach the asymptotic limit of the free theory with the curves approaching the value of 1. In the figure, we see the same behavior with curves of higher temperature being above the lower temperature. We also observe that the curves seem to plateau at larger $n_z$ which is also expected as at large $n_z$ only the highest order coefficient of the polynomial will contribute, settling at a constant value. The upward deviation for $n_z>29$ values is due to the boundary effects. To get rid of this boundary effect, we will fit our data to a conservative maximum bound of $n_z^{max}=25$. 
\par Figure \ref{fig:finite_temp1} (right) plots the value of lattice screening mass at $T=2.90$ GeV obtained from fitting the correlator to the staggered ansatz \eqref{eq:stag_corr}. The correlator is fitted to the ansatz represented by (1,0) and (2,0) which correspond to the number of states considered in the fitting the correlator, {\it i.e.}, 1 and 2 non-oscillating states respectively, with 0 oscillating states. The plot is obtained by keeping the $n_z^{max}=31$ fixed and varying the $n_z^{min}$. The best fit ansatz is chosen by the Akaike criteria AICc \cite{bazavov2019meson}. We see that the first state has a significant contribution atleast till $n_z=20$ and thus, they need to be accounted for when fitting $\Gamma$ and $\Sigma$ as seen in \eqref{eq:Gamma_excited} and \eqref{eq:Sigma_excited} respectively.
\par
\begin{figure}[h]
\centering
\includegraphics[width=0.45\textwidth]{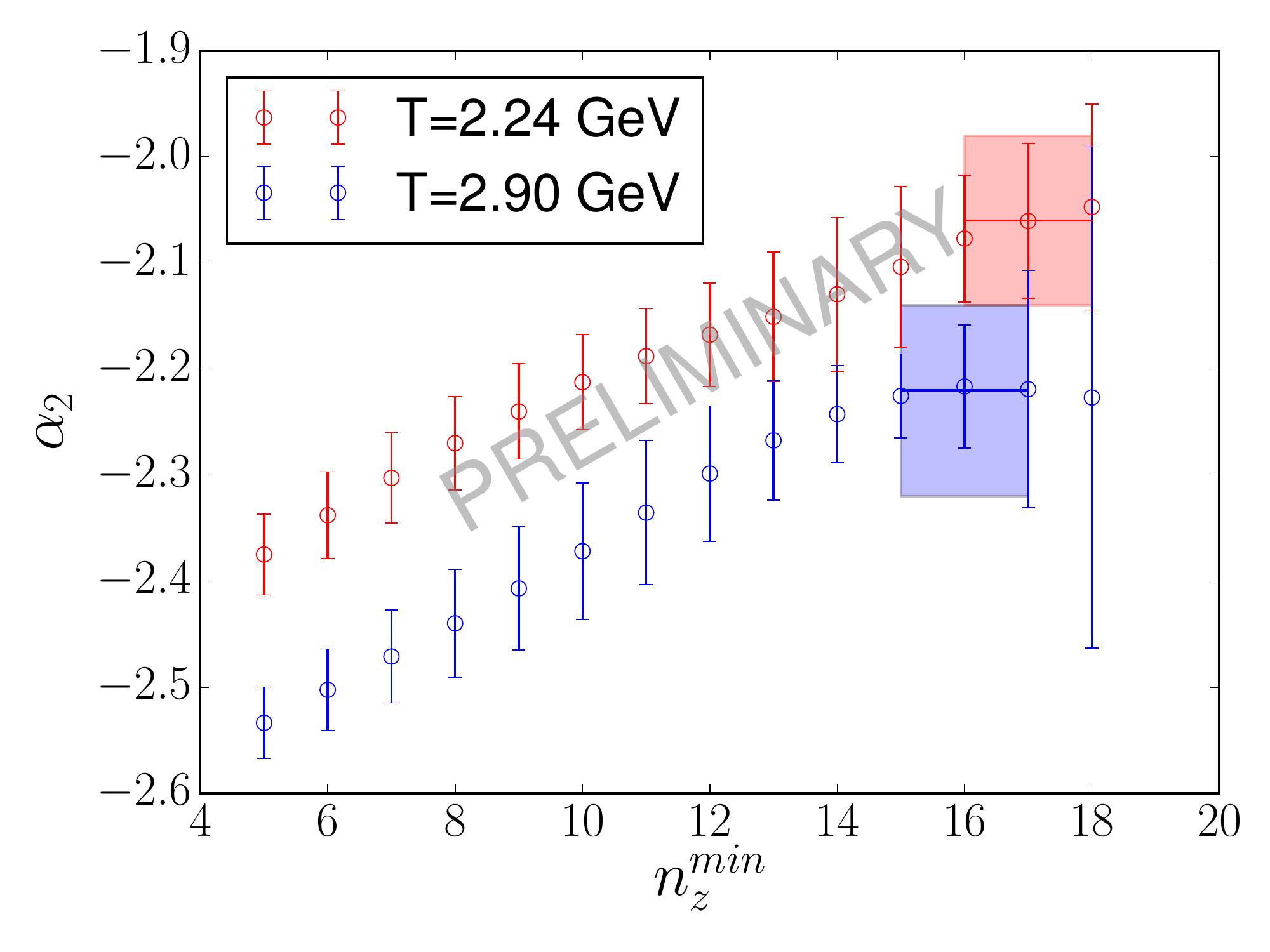}
\includegraphics[width=0.45\textwidth]{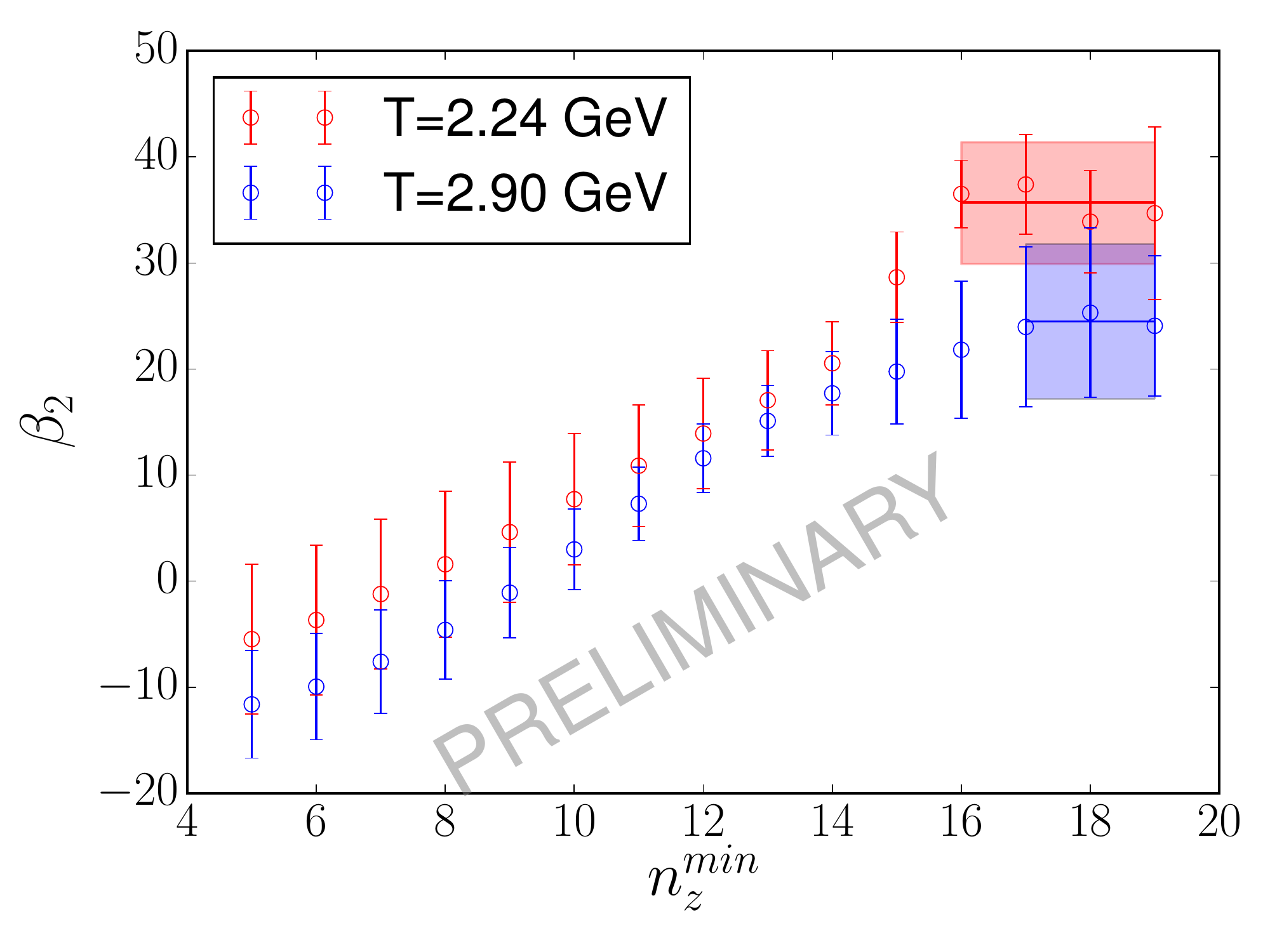}
	\caption{Fit parameters (left) $\alpha_2$ and (right) $\beta_3$ plotted against $n_z^{min}$. The $n^{max} _z=25$ has been fixed for all fit windows. The data points in both the figures are obtained using lattices with dimension $64^3\times8$ at $T=2.24$ GeV and $T=2.90$ GeV using point source and $m_s/m_l=20$. The value and error on the parameter is obtained by bootstrapping over the plateau interval as seen above.}
	\label{fig:finite_temp2}
\end{figure}

 Using \eqref{eq:Gamma_excited} and \eqref{eq:Sigma_excited}, we fit the data obtained on the lattice to obtain the fit coefficients. In figure \ref{fig:finite_temp2}, we have plotted a sample plot for the fit coefficient (left) $\alpha_2$ and (right) $\beta_3$ for $T=2.24$ GeV and  $T=2.90$ GeV. The fit coefficients were obtained by fitting the data in a window such that the contribution of boundary effect as well as the contribution of second excited states or higher is reduced by fixing the $n_z^{max}=25$ while varying the $n_z^{min}$. In the figure, we observe that the value of $\alpha_2$ plateaus as we go to higher $n_z^{min}$ where we get a larger contribution from the ground state and the contribution of the first excited state is expected to exponentially decay. The value of the parameter is taken bootstrapping over the plateau interval. The value interval considered for the plot along with its error is also plotted in the figures.
\par Using the fitting procedure mentioned above, we obtain the table \ref{data} where we have tabulated the fitting coefficients $\alpha_2$, $\alpha_1$, $\beta_4$, and $\beta_3$. The values obtained are quite different from the free theory value while they seem to approach the free theory value with increasing temperature. Using \eqref{eq:mi1} and \eqref{eq:mr2}, we have also tabulated the values of $M_I'$ and $TM_R''$. The value of $M_I'$ also approaches the free theory value with increasing temperature. The value of $M_R''$ while is zero within error but is leaning to have a negative value suggesting the mass decreases with increasing isoscalar chemical potential.
\begin{table}[h]
						\begin{center}
									\begin{tabular}{|c|c|c|c|c|c|c|} \hline

								Temperature $T$ & $\alpha_2$ & $\alpha_1$ & $\beta_4$ & $\beta_3$ & $TM_R''$ & $M_I'$\\ \hline
								2.24 GeV& -2.06(11) & -11.7(5.6) & 6.04(24) & 35.7(5.7) & -0.22(39)&1.43(3)\\							
								2.90 GeV& -2.23(11) & -9.3(6.5) & 6.73(24) & 24.5(7.3) & -0.24(43)&1.49(4)\\\hline
								Free theory & -4 & $10.2$ & 16 & -$81.5$ & 0 & 2\\							
								\hline 
							\end{tabular}
						\end{center}
						\caption{Values for the polynomial fit parameters $\alpha_2$, $\alpha_1$, $\beta_4$ and $\beta_3$ along with the $TM_R''$ and $M_I'$ for two temperatures and free theory. The analysis is done on lattices with $N_\tau=8$ and $N_\sigma=64$.}
						\label{data}
					\end{table}
\section{Conclusion}
In this work, we tried to look at the response of the correlator with the isoscalar chemical potential and obtain correction to the screening mass. We find that the free theory correlator has an oscillating behavior as the screening mass is complex. We verified the free theory expression for the screening correlator derived analytically at finite isoscalar chemical potential by looking at its derivatives on the lattice. Using the symmetric arguments, we extended the analysis to finite temperatures where we obtained the correction to the real part of the screening mass $M_R''$ for two high temperatures. The value of the correction was zero within errors due to large statistical errors but was leaning on the negative side. We also obtained the imaginary part of the screening mass $M_I'$ which seemed to approach the correct free theory limit.

\acknowledgments
This research used the GPU cluster of the Centre for High Energy at the Indian Institute of Science, Bangalore for the generation and analysis of the lattices. The Bielefeld GPU code was used in order to generate the lattices. We also thank Prof. Frithjof Karsch and Dr. Anirban Lahiri for their invaluable input.


\end{document}